\documentclass[12pt]{article}
\usepackage[a4paper,margin=1in]{geometry}
\usepackage{amsmath,amssymb,bm}
\usepackage{siunitx}
\usepackage{booktabs}
\usepackage{graphicx}
\usepackage{caption,subcaption}
\usepackage{float}
\usepackage[utf8]{inputenc}
\usepackage[super,sort&compress]{natbib}
\usepackage[T1]{fontenc}
\usepackage{lmodern}
\usepackage{textcomp}
\usepackage{hyperref}
\usepackage{setspace}
\title {Fragility from Enthalpy Fluctuations and Fictive-Temperature Dynamics}
\author{Biman Bagchi\\
Solid State and Structural Chemistry Unit\\
Indian Institute of Science, Bengaluru 560012, India\\
\texttt{profbiman@gmail.com}}

\date{}

\begin{document}
\maketitle

\begin{abstract}

Fragility and fictive temperature are two central measures of glass formation, but their connection is usually expressed indirectly through structural relaxation times or cooling-rate dependence. Here we formulate this connection in terms of configurational enthalpy fluctuations. The time evolution of the fictive temperature during cooling or aging is controlled by the enthalpy--enthalpy correlation function that also determines the frequency-dependent configurational heat capacity. Fragility enters through the strong temperature dependence of the structural relaxation time and therefore controls how rapidly the corresponding enthalpy-relaxation spectrum shifts to long times near the glass transition. This leads to a direct time-domain description of the onset and growth of the fictive-temperature lag, including the Debye limit, and connects calorimetric relaxation, fragility, and nonequilibrium structural arrest within a common framework.

\end{abstract}

\section{Introduction}

The concept of fictive temperature $T_f$ was introduced to characterize the
nonequilibrium structural state of a glass that falls out of equilibrium during
cooling \cite{tool1946,narayanaswamy1971,moynihan1976}.
Physically, the fictive temperature $T_f$ may be viewed as the temperature at which the
instantaneous structural state of the system would be in equilibrium.
In this sense, $T_f$ is a notional, or effective, temperature introduced to characterize
a nonequilibrium structural state that evolves with time.

While the bath temperature $T(t)$ is externally controlled, the configurational
degrees of freedom relax much more slowly and may fail to keep up with $T(t)$.
As a result, during cooling the structure can become effectively frozen at a
higher temperature, $T_f(t)>T(t)$. The difference $T_f-T$ therefore provides a
measure of the departure from equilibrium. Within this picture, the time evolution
of $T_f(t)$ reflects the progressive arrest of configurational motion as the
structural relaxation time grows rapidly on approaching the glass transition.

A quantitative description of this nonequilibrium behavior was developed by
Tool, Narayanaswamy, and Moynihan, who formulated the kinetics of structural
relaxation in terms of the fictive temperature
\cite{tool1946,narayanaswamy1971,moynihan1976}. In this framework, now known as
the Tool--Narayanaswamy--Moynihan (TNM) formalism, the relaxation time depends
not only on the bath temperature but also on the instantaneous structural state,
typically represented through $T_f$. This phenomenological approach has proven
remarkably successful in describing calorimetric measurements and remains a
standard framework for analyzing aging and cooling processes in glass-forming
systems.

More recent studies have further emphasized the importance of fictive
temperature and structural relaxation in aging, vitrification, and
nonequilibrium dynamics of glasses and polymers
\cite{Mauro2009,Cangialosi2019,Hecksher2015,Cangialosi2024}.
These works highlight the continuing importance of fictive-temperature-based
descriptions and the existence of common structural relaxation timescales
governing different experimental probes. The dependence of the frozen fictive
temperature on cooling rate has also been analyzed by Yue, while the fragility
parameter was introduced by Angell as a measure of the steepness of the
structural relaxation time near $T_g$
\cite{yue2004,angell1991,Richert2002}. Thermodynamic relaxation can also be
characterized through the frequency-dependent heat-capacity spectrum, as
discussed elsewhere \cite{SOB2013}.

In parallel, the fragility parameter $m$, introduced by Angell \cite{angell1991},
quantifies the steepness of the temperature dependence of the structural relaxation
time near the glass transition temperature $T_g$,
\begin{equation}
m = \left. \frac{d \log_{10} \tau}{d (T_g/T)} \right|_{T_g}.
\end{equation}
Fragility distinguishes strong and fragile glass formers and has emerged as a key
descriptor of glassy dynamics, with important implications for viscosity,
relaxation spectra, and thermodynamic anomalies
\cite{angell1995,ediger1996,Ediger2000,ngai2011,Nagel1998}.

A central question, therefore, is how the nonequilibrium evolution of the fictive
temperature is connected to fragility. Experimentally, this connection is probed
through calorimetric measurements, where the dependence of $T_f$ on cooling rate
provides information about the underlying relaxation kinetics
\cite{moynihan1976,roland2005,spieckermann2023}. It is well established that faster
cooling leads to higher frozen fictive temperatures, reflecting incomplete
structural relaxation. Moreover, fragile liquids generally exhibit a more rapid
departure from equilibrium than strong glass formers, indicating an intimate
connection between fragility and fictive-temperature dynamics.

The standard theoretical framework for describing $T_f(t)$ is the
Tool--Narayanaswamy--Moynihan (TNM) formalism
\cite{narayanaswamy1971,moynihan1976}. In this approach, the evolution of the
fictive temperature is governed by a relaxation time that depends not only on
the bath temperature $T$ but also on the structural state, typically represented
through $T_f$. This dependence is often parameterized through a nonlinear mixing
of $T$ and $T_f$, characterized by a phenomenological nonlinearity parameter $x$.
While the TNM framework has been highly successful in describing calorimetric
data, the microscopic interpretation of this nonlinearity remains an important
problem.

%======================================================================

Extensions based on the Adam--Gibbs relation
\cite{AdamGibbs1965,hodge1997} and related thermodynamic approaches have
provided further insight into the relation between relaxation dynamics,
thermodynamics, and nonequilibrium structural evolution. In addition, real
glass-forming systems exhibit nonexponential relaxation, often described by a
stretched-exponential (Kohlrausch--Williams--Watts) form,
\begin{equation}
\phi(t) = \exp\left[ - (t/\tau)^{\beta} \right],
\end{equation}
with the stretching exponent $\beta$ and the nonlinearity parameter $x$ showing
systematic correlations with fragility \cite{ngai2011}.

A particularly important microscopic development in this context of fragility and fictive temperature is the random first-order
transition (RFOT) theory  in which structural
relaxation is viewed as an activated, nucleation-like rearrangement of a
finite amorphous region. The free-energy gain from configurational entropy
is opposed by a mismatch penalty with the surrounding glassy matrix,
naturally introducing a cooperative length scale and an activated
relaxation barrier. \cite{Kirkpatrick1989}

RFOT has also been extended to nonequilibrium aging,
where the structural state, and hence the relaxation rate, depends on the
history of the glass and may be described in terms of a fictive temperature
\cite{LubchenkoWolynes2004}. A complementary mechanical picture was
developed recently by Bagchi, where the mismatch cost is formulated in
terms of elastic heterogeneity and non-affine strain relaxation; this
approach recovers the characteristic configurational-entropy dependence
of the activation barrier while providing an explicitly elastic origin for
the interfacial penalty \cite{BagchiElastic2026}. The present work takes a
different but related route, focusing on configurational enthalpy
fluctuations and their connection to the frequency-dependent heat capacity
and fictive-temperature dynamics.

It is therefore useful to seek a more direct relation between fragility and the
time evolution of the fictive temperature. Existing approaches commonly connect
fragility to relaxation times or viscosities and connect $T_f$ to cooling-rate
dependence, calorimetric response, or structural aging. What is less explicit is
a time-domain formulation in which the Angell fragility parameter enters directly
into the relaxation kernel governing the evolution of $T_f(t)$ and is at the same
time connected to equilibrium configurational fluctuations.

The prior literature establishes several important points:
(a) the fictive temperature $T_f$ depends strongly on cooling rate, which is governed by $\tau(T)$;
(b) fragility enters through the temperature dependence of the structural relaxation time and therefore sets the rate at which the structural clock slows down;
(c) TNM parameters such as $x$, as well as stretching parameters such as $\beta$, show systematic correlations with fragility; and
(d) calorimetric measurements provide a direct experimental window on the same slow structural relaxation.

These observations motivate a formulation in which fragility and
fictive-temperature dynamics are expressed within the same time-correlation
framework.

In this work, we develop such a connection by formulating the evolution of the
fictive temperature within a linear-response framework based on the
configurational enthalpy--enthalpy time correlation function. The same correlation
function determines the configurational part of the frequency-dependent heat
capacity and the delayed enthalpy response to an imposed thermal history.
The fictive temperature is then obtained by mapping this nonequilibrium
configurational enthalpy onto the corresponding equilibrium enthalpy.
Within this formulation, fragility enters through the temperature dependence of
the structural relaxation time and therefore controls the accumulation of reduced
time during cooling. This allows us to obtain explicit relations between $m$ and
the time dependence of $T_f(t)$, the onset of fictive-temperature lag, the
cooling-rate dependence of the frozen fictive temperature within an approximate
freezing criterion, and the temperature shift of the frequency-dependent heat
capacity.

The remainder of the paper develops this framework and explores its implications
for glass-forming liquids of varying fragility.

%Developing such a relation is the goal of the present work.
%

%=======================   2   ===================
%

\section{Enthalpy Fluctuations, Frequency-Dependent Heat Capacity and the Fictive Temperature}

We now derive the relation between the frequency-dependent heat capacity and the
time-dependent fictive temperature. It is useful to clarify at the outset the logic of the construction.  The
frequency-dependent heat capacity is a response function of the slow
configurational degrees of freedom and, in general, its spectrum depends on the
instantaneous structural state of the material.  At the same time, this response
function determines how the configurational enthalpy follows an imposed
temperature history.  The fictive temperature is then obtained by mapping this
nonequilibrium configurational enthalpy onto the equilibrium configurational
enthalpy at an effective temperature.  Thus there is no circular definition:
$C_{p,\rm conf}^{*}(\omega)$ determines the delayed enthalpy response, while
$T_f(t)$ is the thermodynamic label assigned to that enthalpy.  Equivalently,
one may regard the mismatch $T_f(t)-T(t)$ as a measure of the unrelaxed
configurational enthalpy, but the more direct route followed below is to compute
the enthalpy response first and then define $T_f(t)$.

The essential point is that the fictive temperature
is not introduced as an independent dynamical variable. Rather, it is defined from the
instantaneous configurational enthalpy. Thus the logical order is
\begin{equation}
C_{p,\rm conf}^*(\omega)
\quad \longrightarrow \quad
H_{\rm conf}(t)
\quad \longrightarrow \quad
T_f(t).
\end{equation}

At constant pressure, the total enthalpy may be separated into fast and slow
contributions,
\begin{equation}
H(t)=H_{\rm fast}[T(t)]+H_{\rm conf}(t),
\label{eq:H_split_new}
\end{equation}
where $H_{\rm fast}$ contains rapidly equilibrating kinetic, vibrational, librational,
and local contributions, while $H_{\rm conf}$ denotes the slowly relaxing
configurational enthalpy. The fast part remains slaved to the instantaneous bath
temperature,
\begin{equation}
H_{\rm fast}(t)=H_{\rm fast}^{\rm eq}[T(t)].
\end{equation}
The configurational part, however, may lag behind equilibrium during cooling or aging.

The corresponding heat capacity may be decomposed as
\begin{equation}
C_p^{\rm eq}(T)
=
C_p^{\rm fast}(T)+C_p^{\rm conf}(T),
\end{equation}
with
\begin{equation}
C_p^{\rm fast}(T)=
\frac{dH_{\rm fast}^{\rm eq}}{dT},
\qquad
C_p^{\rm conf}(T)=
\frac{dH_{\rm conf}^{\rm eq}}{dT}.
\end{equation}
In dynamic calorimetry this separation appears through the high- and low-frequency
limits of the complex heat capacity,
\begin{equation}
C_p^\infty(T)=C_p^{\rm fast}(T),
\qquad
C_p^0(T)=C_p^{\rm eq}(T).
\end{equation}
The configurational heat-capacity increment is therefore
\begin{equation}
\Delta C_p(T)
=
C_p^0(T)-C_p^\infty(T)
=
C_p^{\rm conf}(T).
\label{eq:DeltaCp_new}
\end{equation}
Thus the fast kinetic and vibrational part is not neglected; it is contained in
$C_p^\infty$. The slow part is relevant to glassy relaxation. We now transfer to the frequency dependent description 
and the right hand side of the above equation gives 
\begin{equation}
C_{p,\rm conf}^*(\omega;T)
=
C_p^*(\omega;T)-C_p^\infty(T).
\label{eq:Cpconf_def_new}
\end{equation}

Although $C_{p,\rm conf}^{*}(\omega;T)$ may itself depend on the structural state
of the liquid, its role here is to determine the delayed configurational
enthalpy response to an imposed bath-temperature history. The fictive temperature
is then obtained by mapping this nonequilibrium configurational enthalpy onto
the corresponding equilibrium configurational enthalpy. Thus $T_f(t)$ is not an
independent external field in the present derivation, but a thermodynamic label
assigned to the unrelaxed enthalpy.

We next relate this quantity to the configurational enthalpy response. Consider an
imposed bath-temperature history $T(t)$. In the linear-response regime, the
configurational enthalpy responds to this thermal history with a delay. In the
frequency domain this response may be written as
\begin{equation}
\delta H_{\rm conf}(\omega)
=
C_{p,\rm conf}^*(\omega;T)\,\delta T(\omega),
\label{eq:H_response_freq_new}
\end{equation}
where $\delta T(\omega)$ is the imposed bath-temperature perturbation. Importantly,
the perturbing temperature in Eq.~(\ref{eq:H_response_freq_new}) is the bath
temperature, not $T_f-T$. The fictive temperature will be introduced only after
the configurational enthalpy has been obtained.

Following the dynamic heat-capacity formulation of Saito, Ohmine and Bagchi \cite{SOB2013}, the
configurational part of the heat capacity is related to the normalized
configurational enthalpy--enthalpy time-correlation function (EETCF), $\Phi_H(t;T)$, defined by,
\begin{equation}
\Phi_H(t;T)
=
\frac{
\left\langle \delta H_{\rm conf}(t)\delta H_{\rm conf}(0)\right\rangle_T
}{
\left\langle [\delta H_{\rm conf}(0)]^2\right\rangle_T
}.
\label{eq:PhiH_new}
\end{equation}
Using the above definition, we can now write down the corresponding configurational heat capacity as \cite{SOB2013}
\begin{equation}
C_{p,\rm conf}^*(\omega;T)
=
\Delta C_p(T)
\left[
1-i\omega \hat{\Phi}_H(\omega;T)
\right],
\label{eq:Cp_Phi_new}
\end{equation}
where
\begin{equation}
\hat{\Phi}_H(\omega;T)
=
\int_0^\infty dt\,e^{-i\omega t}\Phi_H(t;T).
\end{equation}
Equation~(\ref{eq:Cp_Phi_new}) contains the microscopic input: the slow heat-capacity
spectrum is determined by the configurational enthalpy fluctuations.

The equivalent time-domain form follows by inverse transformation. Substituting the frequency-domain expression for
$C_{p,\rm conf}^{*}(\omega)$ from Eq.~(14) into the linear-response relation,
Eq.~(12), and then transforming back to the time domain yields a convolution
expression for the configurational enthalpy. Physically, this convolution
represents the delayed response of the configurational degrees of freedom to the
imposed temperature history, with the relaxation kernel determined by the
enthalpy--enthalpy correlation function.
For a general
temperature history $T(t)$, the configurational enthalpy is
\begin{equation}
H_{\rm conf}(t)
=
H_{\rm conf}^{\rm eq}[T(t)]
-
\int_0^t ds\,
\Delta C_p[T(s)]\,
\Phi_H(t-s;T)\,
\dot T(s).
\label{eq:Hconf_convolution_new}
\end{equation}
The first term is the enthalpy that would be obtained if the configurational degrees
of freedom remained in equilibrium with the bath. The second term is the delayed
response arising from the finite relaxation time of the slow structural modes.

We now define the fictive temperature. It is the equilibrium temperature at which
the configurational enthalpy would have the instantaneous nonequilibrium value:
\begin{equation}
H_{\rm conf}(t)
\equiv
H_{\rm conf}^{\rm eq}[T_f(t)].
\label{eq:Tf_definition_new}
\end{equation}
Therefore,
\begin{equation}
H_{\rm conf}^{\rm eq}[T_f(t)]
=
H_{\rm conf}^{\rm eq}[T(t)]
-
\int_0^t ds\,
\Delta C_p[T(s)]\,
\Phi_H(t-s;T)\,
\dot T(s).
\label{eq:Tf_exact_enthalpy_mapping_new}
\end{equation}
This is the direct and desired relation between the configurational enthalpy response and the
fictive temperature.

For small to moderate departures from equilibrium, we linearize
$H_{\rm conf}^{\rm eq}[T_f(t)]$ about $T(t)$:
\begin{equation}
H_{\rm conf}^{\rm eq}[T_f(t)]
\simeq
H_{\rm conf}^{\rm eq}[T(t)]
+
\Delta C_p[T(t)]\,[T_f(t)-T(t)].
\label{eq:H_linearize_new}
\end{equation}
We now substitute the above into Eq.~(\ref{eq:Tf_exact_enthalpy_mapping_new}) gives
\begin{equation}
T_f(t)-T(t)
=
-
\frac{1}{\Delta C_p[T(t)]}
\int_0^t ds\,
\Delta C_p[T(s)]\,
\Phi_H(t-s;T)\,
\dot T(s).
\label{eq:Tf_general_Cp_new}
\end{equation}
If the configurational heat-capacity increment varies only weakly over the
temperature interval relevant to structural arrest, $\Delta C_p$ may be taken
outside the integral, leading to cancellation between numerator and denominator.This simplifies the above expression to
\begin{equation}
T_f(t)
=
T(t)
-
\int_0^t ds\,
\Phi_H(t-s;T)\,\dot T(s).
\label{eq:Tf_simple_convolution_new}
\end{equation}

Equation~(\ref{eq:Tf_simple_convolution_new}) is the desired time-domain relation.
It has been obtained in the order
\[
C_{p,\rm conf}^*(\omega)
\rightarrow
H_{\rm conf}(t)
\rightarrow
T_f(t),
\]
and therefore avoids any circular use of $T_f$ as the initial perturbation.

For a linear cooling protocol,
\begin{equation}
T(t)=T_0-qt,\qquad q>0,
\end{equation}
so that $\dot T=-q$. Equation~(\ref{eq:Tf_simple_convolution_new}) becomes
\begin{equation}
T_f(t)
=
T(t)
+
q\int_0^t ds\,\Phi_H(t-s;T).
\label{eq:Tf_linear_cooling_new}
\end{equation}
This form shows explicitly that the fictive temperature exceeds the bath temperature
during cooling because the configurational enthalpy retains memory of earlier,
higher-temperature states.

In a glass-forming liquid, the structural relaxation time changes strongly with
temperature during cooling. As a result, equal intervals of laboratory time do
not correspond to equal amounts of structural relaxation. At high temperature,
where $\tau(T)$ is short, the structure evolves rapidly, whereas near the glass
transition the same interval of laboratory time produces only a very small amount
of configurational relaxation. It is therefore convenient to introduce a reduced
or material time that measures the accumulated progress of structural relaxation
along the thermal history. Following the Tool--Narayanaswamy description of
aging and later material-time formulations, we define the reduced time by

\begin{equation}
d\zeta = \frac{dt}{\tau[T(t)]},
\end{equation}
so that
\begin{equation}
\zeta(t)=\int_0^t \frac{dt'}{\tau[T(t')]}.
\label{eq:zeta_new}
\end{equation}

The reduced time therefore represents the total amount of configurational
relaxation accumulated up to time $t$. When $\tau(T)$ becomes very large near
the glass transition, the internal clock effectively slows down and $\zeta(t)$
increases only weakly even though laboratory time continues to advance.

The preceding convolution was written for a fixed reference temperature.  During
cooling, however, the relaxation spectrum itself shifts continuously because the
structural relaxation time changes with temperature.  To extend the convolution
to a nonisothermal history, we adopt the standard reduced-time, or material-time,
approximation used in TNM-type descriptions of structural relaxation.  The
assumption is not that the relaxation function is strictly temperature
independent in laboratory time, but rather that its shape is approximately
invariant when expressed in terms of the accumulated internal time $\zeta$.

Thus, the kernel is assumed to depend on the amount of structural relaxation
elapsed between the earlier time $s$ and the observation time $t$, rather than
on the laboratory-time interval $t-s$ alone.  This accumulated relaxation is
measured by

\begin{equation}
\zeta(t)-\zeta(s)=\int_s^t \frac{du}{\tau[T(u)]}.
\end{equation}

Consequently, the nonisothermal relaxation kernel is approximated as
\begin{equation}
\Phi_H(t-s;T)\rightarrow \Phi_H[\zeta(t)-\zeta(s)].
\end{equation}

This step is the usual thermorheological-simplicity assumption: cooling changes
the rate at which the internal clock advances, but does not strongly change the
shape of the normalized configurational enthalpy relaxation function.  If the
spectral shape itself changes appreciably with temperature, the present
expression must be generalized by allowing $\Phi_H$ to depend explicitly on both
temperature and structural state.

The fictive-temperature relation then becomes
\begin{equation}
T_f(t)
=
T(t)
-
\int_0^t ds\,
\Phi_H[\zeta(t)-\zeta(s)]\,\dot T(s),
\label{eq:Tf_reduced_time_new}
\end{equation}
or, for linear cooling,
\begin{equation}
T_f(t)
=
T_0-qt
+
q\int_0^t ds\,
\Phi_H[\zeta(t)-\zeta(s)].
\label{eq:Tf_reduced_linear_new}
\end{equation}

\subsection{Debye limit and emergence of the Tool equation}

Finally, consider the single-relaxation-time, or Debye, limit of the
configurational enthalpy relaxation. In reduced time this corresponds to
\begin{equation}
\Phi_H(\Delta \zeta)=\exp(-\Delta \zeta).
\end{equation}
For a general thermal history, Eq.~(28) may then be written as
\begin{equation}
T_f(t)
=
T(t)
-
\int_0^t ds\,
\exp[-\zeta(t)+\zeta(s)]\,\dot T(s).
\end{equation}
Define the lag variable
\begin{equation}
y(t)=T_f(t)-T(t).
\end{equation}
Then
\begin{equation}
y(t)
=
-
\int_0^t ds\,
\exp[-\zeta(t)+\zeta(s)]\,\dot T(s).
\end{equation}
Differentiating this expression with respect to $t$ gives
\begin{equation}
\frac{dy}{dt}
=
-\dot T(t)
-
\frac{d\zeta}{dt}\,y(t).
\end{equation}
Since
\begin{equation}
\frac{d\zeta}{dt}=\frac{1}{\tau[T(t)]},
\end{equation}
we obtain
\begin{equation}
\frac{dy}{dt}
=
-\dot T(t)-\frac{y(t)}{\tau[T(t)]}.
\end{equation}
Using $y(t)=T_f(t)-T(t)$, so that
\begin{equation}
\frac{dy}{dt}=\frac{dT_f}{dt}-\dot T(t),
\end{equation}
one finally obtains
\begin{equation}
\frac{dT_f}{dt}
=
\frac{T(t)-T_f(t)}{\tau[T(t)]}.
\end{equation}
Thus the usual Tool-type equation appears as the single-relaxation-time limit
of the more general heat-capacity response formulation. The microscopic object
remains the configurational enthalpy--enthalpy time-correlation function, or
equivalently $C_{p,\rm conf}^{*}(\omega)$.

The fragility $m$ enters through the enthalpy-enthalpy time correlation function (EETCF) defined above.

% ====================== 3 ========================
%=====================Section 3 =======================

\section{Fragility and Fictive Temperature Dynamics }

We now connect the fictive-temperature dynamics developed above to the fragility
parameter $m$. The essential point is simple. The shape of the configurational
enthalpy relaxation function is determined by the enthalpy--enthalpy time
correlation function, or equivalently by the configurational part of the
frequency-dependent heat capacity. Fragility enters through the temperature
dependence of the structural relaxation time that sets the clock for this
relaxation.

\subsection{Fragility as the temperature dependence of the structural clock}

The Angell fragility parameter is defined by
\begin{equation}
m
=
\left.
\frac{d\log_{10}\tau(T)}
{d(T_g/T)}
\right|_{T=T_g},
\label{eq:fragility_def_new}
\end{equation}
where $T_g$ is the operational glass transition temperature and $\tau(T)$ is
the structural relaxation time.

Near $T_g$, the most direct local representation consistent with this definition is
\begin{equation}
\log_{10}\tau(T)
=
\log_{10}\tau_g
+
m\left(\frac{T_g}{T}-1\right),
\label{eq:tau_local_m_new}
\end{equation}
where
\begin{equation}
\tau_g\equiv \tau(T_g).
\end{equation}
Thus,
\begin{equation}
\tau(T)
=
\tau_g\,10^{m\left(\frac{T_g}{T}-1\right)}.
\label{eq:tau_m_new}
\end{equation}

Equation~(\ref{eq:tau_m_new}) should not be viewed as a new global empirical
law for the relaxation time. It is a local representation near $T_g$ that
inserts the experimentally defined fragility into the fictive-temperature
dynamics. A VFT, parabolic, or Adam--Gibbs form may be used instead, provided
that its slope at $T_g$ gives the same value of $m$.

\subsection{Reduced time and the general relation for $T_f(t)$}

From the dynamic heat-capacity formulation of the previous section, the
fictive-temperature trajectory may be written as
\begin{equation}
T_f(t)
=
T(t)
-
\int_0^t ds\,
\Phi_H[\zeta(t)-\zeta(s)]\,\dot T(s),
\label{eq:Tf_general_section3}
\end{equation}
where $\Phi_H$ is the normalized configurational enthalpy relaxation function.
The reduced time is
\begin{equation}
\zeta(t)
=
\int_0^t \frac{dt'}{\tau[T(t')]}.
\label{eq:zeta_section3}
\end{equation}
For clarity, we first write the result using a bath-temperature-dependent
relaxation time. Additional dependence on $T_f$ may be included in a TNM-like
extension, but is not needed for the basic relation between fragility and
fictive-temperature evolution.

For a linear cooling protocol,
\begin{equation}
T(t)=T_0-qt,\qquad q>0,
\label{eq:linear_cooling_section3}
\end{equation}
one has $\dot T=-q$, and Eq.~(\ref{eq:Tf_general_section3}) becomes
\begin{equation}
T_f(t)
=
T_0-qt
+
q\int_0^t ds\,
\Phi_H[\zeta(t)-\zeta(s)].
\label{eq:Tf_linear_section3}
\end{equation}

Substituting Eq.~(\ref{eq:tau_m_new}) into Eq.~(\ref{eq:zeta_section3}) gives
\begin{equation}
\zeta(t)-\zeta(s)
=
\int_s^t
\frac{du}
{\tau_g\,10^{m\left(\frac{T_g}{T_0-qu}-1\right)}}.
\label{eq:zeta_m_section3}
\end{equation}
Therefore,
\begin{equation}
T_f(t)
=
T_0-qt
+
q\int_0^t ds\,
\Phi_H\!\left[
\int_s^t
\frac{du}
{\tau_g\,10^{m\left(\frac{T_g}{T_0-qu}-1\right)}}
\right].
\label{eq:Tf_m_general_section3}
\end{equation}

Equation~(\ref{eq:Tf_m_general_section3}) is the central time-domain result of
this section. It shows that fragility affects $T_f(t)$ by controlling the rate at
which reduced time accumulates during cooling. The microscopic relaxation
function $\Phi_H$ is supplied by the configurational enthalpy TCF, while $m$
sets the temperature dependence of the structural clock.

\subsection{Debye limit}

In the Debye limit, the heat-capacity formulation reduces to the local Tool-type
equation derived in Sec.~2.1. Substituting the fragility representation of
$\tau(T)$ into that equation gives
\begin{equation}
\frac{dT_f}{dt}
=
\frac{T(t)-T_f(t)}
{\tau_g\,10^{m(T_g/T(t)-1)}} .
\label{eq:Debye_m_section3}
\end{equation}

This equation is useful because it displays the role of fragility transparently.
For a fragile liquid, $\tau(T)$ grows rapidly as $T$ is lowered through the glass
transition region. The denominator in Eq.~(\ref{eq:Debye_m_section3}) therefore
increases rapidly, and $T_f(t)$ ceases to follow the bath temperature. For a
strong liquid, the growth of $\tau(T)$ is weaker, and the fictive temperature
remains closer to equilibrium over a broader temperature range.

\subsection{Near-$T_g$ form}

Near $T_g$,
\begin{equation}
\frac{T_g}{T}-1
\simeq
\frac{T_g-T}{T_g}.
\end{equation}
Equation~(\ref{eq:tau_m_new}) then becomes
\begin{equation}
\tau(T)
\simeq
\tau_g\exp[\mu(T_g-T)],
\qquad
\mu=\frac{m\ln 10}{T_g}.
\label{eq:tau_nearTg_section3}
\end{equation}

The parameter $\mu$ is the local thermal slope of the structural clock. It is
directly proportional to fragility. Thus a large value of $m$ implies that the
enthalpy relaxation spectrum shifts rapidly to low frequency as the temperature
is lowered, causing earlier arrest of the configurational enthalpy and a larger
fictive-temperature lag.

\subsection{Physical interpretation}

The result of this section can be summarized as follows. The configurational
enthalpy TCF determines the relaxation function $\Phi_H$. The frequency-dependent
heat capacity measures the same relaxation spectrum. The fictive temperature is
obtained by mapping the slow configurational enthalpy lag onto an equivalent
equilibrium enthalpy. Fragility enters through the temperature dependence of the
relaxation time or, equivalently, through the temperature shift of the
calorimetric loss spectrum.

Thus,
\begin{equation}
\text{enthalpy TCF}
\quad \longrightarrow \quad
C_{p,\rm conf}^*(\omega)
\quad \longrightarrow \quad
T_f(t),
\end{equation}
while
\begin{equation}
m
\quad \longrightarrow \quad
\tau(T)
\quad \longrightarrow \quad
\zeta(t)
\quad \longrightarrow \quad
T_f(t).
\end{equation}

This is the desired connection between fragility and fictive-temperature dynamics.
Fragility does not change the basic microscopic origin of the response; it controls
 how rapidly the enthalpy relaxation spectrum shifts with temperature.

% ===================== Section 4 =============

\section{Onset and Freezing of the Fictive Temperature}

The preceding formulation allows a robust statement about the onset of
nonequilibrium behavior during cooling. The Angell fragility parameter \(m\)
is defined from the equilibrium, or metastable-equilibrium, supercooled liquid
as \(T_g\) is approached from above. It therefore controls the temperature
dependence of the structural relaxation time in the regime where the system is
still able to equilibrate.

During cooling, the fictive temperature follows the bath temperature as long as
the accumulated reduced time is large enough for configurational equilibration.
Departure from equilibrium begins when the structural relaxation time becomes
comparable to the timescale over which the bath temperature changes appreciably.
Thus the condition for the onset of a fictive-temperature lag may be written
schematically as
\[
\tau(T_{\rm on}) \sim \frac{\Delta T}{q},
\]
where \(q=|dT/dt|\) is the cooling rate and \(\Delta T\) is the temperature
interval over which the relaxation time changes significantly. This criterion
should be viewed as an onset condition, not as a precise prediction of the final
frozen fictive temperature.

Using the local fragility representation near \(T_g\),
\[
\tau(T)\simeq \tau_g
10^{m(T_g/T-1)},
\]
one sees that a larger value of \(m\) produces a sharper increase of
\(\tau(T)\) upon cooling. Consequently, fragile liquids develop a
fictive-temperature lag over a narrower temperature interval, whereas strong
liquids remain closer to equilibrium over a broader interval.

Below \(T_g\), however, the system is no longer in equilibrium at the bath
temperature. The relaxation kernel governing further evolution of \(T_f(t)\)
cannot in general be identified with the equilibrium enthalpy--enthalpy
correlation function at \(T\). A quantitative prediction of the final frozen
fictive temperature therefore requires a nonequilibrium continuation of the
structural clock, for example through a TNM-like relaxation time
\(\tau(T,T_f)\). The present theory should therefore be interpreted as giving
a microscopic basis for the onset and growth of the fictive-temperature lag,
rather than a fully quantitative formula for the final frozen value of
\(T_f\).

%=========================================

\subsection{Definition of frozen fictive temperature}

The frozen fictive temperature is defined as the asymptotic value of $T_f(t)$
after the system has fallen out of equilibrium during cooling,
\begin{equation}
T_f^{\rm fr}(q) \equiv \lim_{t \to \infty} T_f(t).
\end{equation}
Physically, it corresponds to the temperature at which structural relaxation
effectively ceases on the timescale set by the cooling rate.

%\subsection{General memory-function expression}

Starting from the time-domain relation derived in Sec.~2, the frozen fictive
temperature is defined and controlled by the point at which the reduced time no longer
accumulates appreciably during cooling.

Hence the frozen fictive temperature is determined by the region where relaxation
effectively stops, i.e., where reduced time ceases to accumulate.

\subsection{Physical freezing criterion}

A physically transparent approximation is obtained by noting that freezing occurs
when the structural relaxation time becomes comparable to the inverse cooling rate,
\begin{equation}
\tau(T_f^{\rm fr}) \sim \frac{\Delta T}{q},
%\tau(T_f^{\rm fr}) \sim \frac{1}{q},
\label{eq:freezing_condition}
\end{equation}
where $\Delta T$ is the temperature over which the system must relax.
This criterion follows directly from the memory formalism: when
$\tau(T)$ grows rapidly, the accumulation of reduced time
\[
\zeta = \int dt/\tau(T)
\]
becomes negligible, so the kernel $\Phi$ no longer evolves and the structure freezes.

\subsection{Relation between $T_f^{\rm fr}$ and fragility}

Using the fragility-based expression for the relaxation time,
\begin{equation}
\tau(T)
=
\tau_g\,10^{\,m\left(\frac{T_g}{T}-1\right)},
\end{equation}
and substituting into the freezing condition Eq.~(\ref{eq:freezing_condition}),
we obtain
\begin{equation}
\tau_g\,10^{\,m\left(\frac{T_g}{T_f^{\rm fr}}-1\right)}
=
\frac{1}{q}.
\end{equation}

Taking $\log_{10}$,
\begin{equation}
\log_{10}\left(\frac{1}{q\tau_g}\right)
=
m\left(\frac{T_g}{T_f^{\rm fr}}-1\right).
\end{equation}

Solving for $T_f^{\rm fr}$ gives the central result,
\begin{equation}
T_f^{\rm fr}(q)
=
\frac{T_g}{1+\frac{1}{m}\log_{10}\!\left(\frac{\Delta T}{q\tau_g}\right)}
.
\label{eq:Tf_fr_main}
\end{equation}

This is a closed-form expression relating the frozen fictive temperature,
cooling rate, and fragility.

\subsection{Near-$T_g$ linearized form}

For small deviations from $T_g$, we expand
\begin{equation}
\frac{T_g}{T_f^{\rm fr}} - 1
\simeq \frac{T_g - T_f^{\rm fr}}{T_g}.
\end{equation}
Substituting into the logarithmic relation yields
\begin{equation}
T_g - T_f^{\rm fr}
\simeq
\frac{T_g}{m}\log_{10}\!\left(\frac{1}{q\tau_g}\right).
\end{equation}

Thus, we obtain,
\begin{equation}
T_f^{\rm fr}(q)
\simeq
T_g
-
\frac{T_g}{m}\log_{10}\!\left(\frac{1}{q\tau_g}\right)
.
\label{eq:Tf_fr_linear}
\end{equation}

This shows that the shift of the fictive temperature with cooling rate is
inversely proportional to fragility.

\subsection{Connection to experiments}

Equations~(\ref{eq:Tf_fr_main}) and (\ref{eq:Tf_fr_linear}) provide a direct
connection to calorimetric measurements:

\begin{itemize}
\item A plot of $T_f^{\rm fr}$ versus $\log q$ yields a slope proportional to $1/m$.
\item Fragile liquids (large $m$) show weaker dependence of $T_f^{\rm fr}$ on $q$.
\item Strong liquids (small $m$) show a larger shift of $T_f^{\rm fr}$ with cooling rate.
\end{itemize}

Thus, the frozen fictive temperature serves as a direct experimental probe of fragility.

\subsection{Microscopic interpretation}

Within the present framework, the origin of this relation is clear.

The structural relaxation function $\Phi$ is determined by the configurational
enthalpy--enthalpy correlation function.
Fragility does not modify the shape of $\Phi$, but controls the rate at which
reduced time accumulates via $\tau(T)$.

A larger fragility implies a sharper growth of $\tau(T)$ below $T_g$,
leading to earlier arrest of the dynamics and hence a higher frozen fictive temperature.
The result provides a direct connection between nonequilibrium glass formation
and equilibrium dynamic properties encoded in fragility.
%
%
% =======================  5  ======================
%
\section{Calorimetric Loss Spectrum and Fragility}

The formulation developed above shows that the fictive-temperature dynamics and
the calorimetric spectrum originate from the same configurational enthalpy
relaxation. We now use this connection to obtain a frequency-domain expression
for the fragility parameter.

The configurational part of the complex heat capacity may be written as
\begin{equation}
C_p^*(\omega;T)
=
C_p^\infty(T)+\Delta C_p(T)\chi(\omega;T),
\label{eq:Cpstar_chi}
\end{equation}
where $C_p^\infty$ is the fast high-frequency contribution, $\Delta C_p$ is the
configurational heat-capacity increment, and $\chi(\omega;T)$ is the normalized
structural susceptibility. If $\Phi(t;T)$ is the normalized configurational
enthalpy relaxation function, then
\begin{equation}
\chi(\omega;T)
=
1-i\omega \hat{\Phi}(\omega;T),
\label{eq:chi_phi}
\end{equation}
and hence
\begin{equation}
C_p''(\omega;T)
=
\Delta C_p(T)\,\omega
\int_0^\infty \Phi(t;T)\sin(\omega t)\,dt .
\label{eq:Cpdoubleprime_general}
\end{equation}
Thus the loss spectrum of the frequency-dependent heat capacity directly probes
the same configurational enthalpy relaxation function that controls the
fictive-temperature response.

Under thermorheological simplicity, the relaxation function depends on time only
through the scaled variable $t/\tau(T)$,
\begin{equation}
\Phi(t;T)=\varphi\!\left(\frac{t}{\tau(T)}\right),
\label{eq:phi_scaling}
\end{equation}
where $\varphi$ is a temperature-independent master function. Substitution into
Eq.~(\ref{eq:Cpdoubleprime_general}) gives
\begin{equation}
C_p''(\omega;T)
=
\Delta C_p(T)\,
{\cal F}[\omega\tau(T)],
\label{eq:Cp_scaling}
\end{equation}
with
\begin{equation}
{\cal F}(y)
=
y\int_0^\infty \varphi(x)\sin(yx)\,dx .
\label{eq:F_def}
\end{equation}
Therefore, the temperature dependence of the calorimetric loss peak is governed
by the single product $\omega\tau(T)$.

Let $\omega_p(T)$ denote the frequency at which $C_p''(\omega;T)$ is maximum.
The peak condition gives
\begin{equation}
\omega_p(T)\tau(T)=y_p,
\label{eq:peak_master}
\end{equation}
where $y_p$ is a dimensionless number determined only by the shape of the
relaxation spectrum. For Debye relaxation, $y_p=1$; for nonexponential
relaxation, $y_p$ differs from unity but remains independent of temperature if
the spectral shape is invariant.

Using the local fragility representation
\begin{equation}
\tau(T)
=
\tau_g\,10^{m\left(\frac{T_g}{T}-1\right)},
\label{eq:tau_fragility_cp}
\end{equation}
we obtain
\begin{equation}
\omega_p(T)
=
\frac{y_p}{\tau_g}
10^{-m\left(\frac{T_g}{T}-1\right)} .
\label{eq:omega_peak_m}
\end{equation}
Equivalently,
\begin{equation}
\log_{10}\omega_p(T)
=
\log_{10}\!\left(\frac{y_p}{\tau_g}\right)
-
m\left(\frac{T_g}{T}-1\right).
\label{eq:log_omega_peak_m}
\end{equation}
Thus the slope of $\log_{10}\omega_p$ versus $T_g/T$ is $-m$.

This leads to the frequency-domain definition
\begin{equation}
m
=
-
\left.
\frac{d\log_{10}\omega_p}
{d(T_g/T)}
\right|_{T=T_g}.
\label{eq:m_from_omega_peak}
\end{equation}
This expression is equivalent to the usual Angell definition in terms of
$\tau(T)$, but it has a different physical interpretation: fragility is encoded
in the temperature shift of the enthalpy-fluctuation spectrum.

\begin{figure}[H]
\centering
\includegraphics[width=0.9\linewidth]{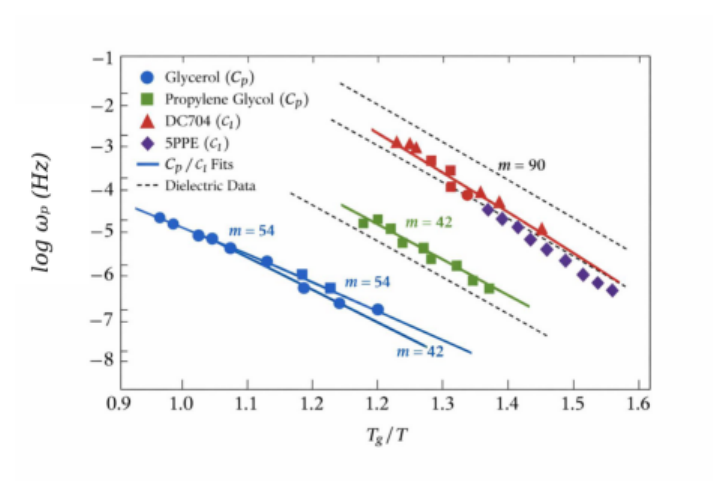}
\caption{
Logarithm of the calorimetric loss-peak frequency $\omega_p$ as a function of
inverse reduced temperature $T_g/T$ for representative glass-forming liquids.
Data for glycerol and propylene glycol are taken from specific-heat spectroscopy
measurements by Birge and Nagel, where $\omega_p(T)$ is obtained from the peak
of $C_p''(\omega)$. Data for DC704 and 5PPE are taken from longitudinal
specific-heat spectroscopy measurements reported by Jakobsen et al.\ and Roed
et al., where $\omega_p$ is extracted from the imaginary part of the longitudinal
specific heat. In all cases, the peak frequency is plotted against $T_g/T$ using
literature values of $T_g$. The solid lines represent fits to Eq.~(72),
$\log_{10}\omega_p(T)=\log_{10}(y_p/\tau_g)-m(T_g/T-1)$,
where $\tau_g$ is the structural relaxation time at the glass transition
temperature $T_g$, and $m$ is the fragility parameter.
The slopes of these lines directly yield the fragility parameter $m$. Dashed lines
indicate representative dielectric-relaxation trends, illustrating that different
experimental probes (calorimetric, dielectric, mechanical) exhibit identical
temperature dependence but differ by a constant prefactor. This demonstrates the
existence of a common structural relaxation clock governing all linear-response
functions, while the absolute timescale depends on the specific observable.
}
\label{fig:fragility_cp}
\end{figure}

Experimental evidence supports this interpretation. Specific-heat spectroscopy
measurements by Birge and Nagel on glycerol and propylene glycol showed that the
calorimetric loss peak follows the same temperature dependence as structural
relaxation times \cite{BirgeNagel1986}. More recent measurements by Jakobsen
et al.\ demonstrated that dielectric, mechanical, and longitudinal specific-heat
loss-peak frequencies exhibit the same temperature dependence over many decades,
differing only by constant prefactors \cite{Jakobsen2012}. Roed et al. further
showed that this proportionality persists under changes of temperature and
pressure in 5PPE \cite{Roed2015}. 
These observations indicate that different
linear-response probes share a common structural clock. This conclusion is also 
consistent with the ``inner-clock'' perspective
developed by Peredo-ortiz et al., who emphasized that different linear-response
functions in glass-forming liquids are governed by a common structural relaxation
timescale.\cite {PeredoOrtiz2022InnerClocks, Dyre2024}
%===========================================
%===========================================
Figure~\ref{fig:fragility_cp} illustrates this result using representative
calorimetric and longitudinal specific-heat data. The approximately linear
dependence of $\log_{10}\omega_p$ on $T_g/T$ near $T_g$ is the central prediction
of Eq.~(\ref{eq:log_omega_peak_m}). The slopes give the fragility parameter,
while vertical offsets reflect probe-dependent prefactors rather than different
temperature dependencies.

% KEEP YOUR EXISTING FIGURE 1 BLOCK AND CAPTION EXACTLY HERE.

The data in Fig.~\ref{fig:fragility_cp} therefore demonstrate that fragility can
be obtained directly from calorimetric spectroscopy. This provides a
fluctuation-based alternative to the traditional kinetic definition and connects
fragility to the spectral evolution of configurational enthalpy fluctuations.

% ========================= \textbf{ Figure 1 } %============================
%\clearpage

% ======================== Figure 1 ends %========================

The data shown in Fig.~1 are constructed from published calorimetric and
longitudinal specific-heat spectroscopy measurements. For glycerol and
propylene glycol, we use the specific-heat spectroscopy data of Birge and
Nagel \cite{BirgeNagel1986}, where the loss peak in $C_p''(\omega)$ directly
provides the characteristic frequency $\omega_p(T)$. For the van der Waals
liquids DC704 and 5PPE, we use longitudinal specific-heat spectroscopy
results reported by Jakobsen et al.\ \cite{Jakobsen2012} and Roed et al.\
\cite{Roed2015}, where the peak frequency is extracted from the imaginary
part of the longitudinal specific heat. In each case, the peak frequencies
were compiled as a function of temperature and plotted against the inverse
reduced temperature $T_g/T$. The approximately linear behavior observed
near $T_g$ confirms Eq.~(72), with slopes that directly yield the fragility
parameter $m$. The small differences in absolute magnitude between systems
and between calorimetric and dielectric probes reflect probe-dependent
prefactors, while the identical slopes demonstrate the existence of a
common structural relaxation clock.

The figure demonstrates that fragility can be extracted directly from calorimetric data, providing a fluctuation-based alternative to the traditional kinetic definition. This establishes a direct connection between fragility and the spectral properties of enthalpy fluctuations, thereby bridging thermodynamics, linear response, and glassy dynamics.

% ============================  6 ================

\section{Strong and Fragile Liquids: Numerical Illustration}

We now illustrate the consequences of the above theory for the fictive-temperature
trajectory during cooling. The purpose of this calculation is not to fit a
specific experiment, but to show how the contrast between strong and fragile
glass formers appears in $T_f(T)$.

We use representative material parameters for silica and ethanol. Silica is a
canonical strong glass former with low fragility, while ethanol is a fragile
molecular liquid. The values of $T_g$ and $m$ are taken from standard
compilations of glass-forming liquids \cite{angell1995,ediger1996}. The
relaxation time at the glass transition is taken to be $\tau_g=100$ s, the
usual calorimetric convention.

% KEEP YOUR EXISTING TABLE 1 EXACTLY HERE.

The relaxation time is represented locally near $T_g$ as
\begin{equation}
\tau(T)
=
\tau_g\,10^{m\left(\frac{T_g}{T}-1\right)}.
\label{eq:tau_numeric}
\end{equation}

The fictive temperature is obtained by numerically integrating the Debye-limit
equation derived in Sec.~2.1, using the relaxation time in Eq.~(73).

under the linear cooling protocol
\begin{equation}
T(t)=T_0-qt,
\qquad q=2~{\rm K/min}.
\end{equation}
Both $T$ and $T_f$ are scaled by $T_g$ so that systems with very different glass
transition temperatures can be compared on the same plot.

% KEEP YOUR EXISTING FIGURE 2 BLOCK AND CAPTION EXACTLY HERE.

\begin{figure}[H]
    
\centering
\includegraphics[width=0.8\linewidth]{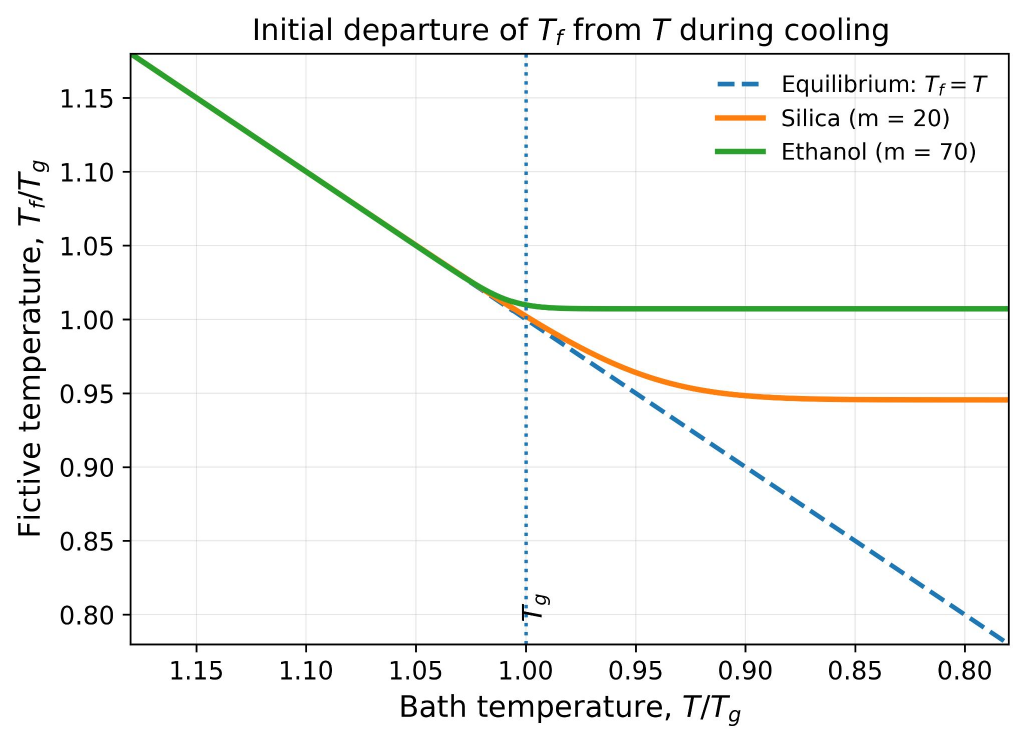}
\caption{
Comparison of fictive-temperature evolution during linear cooling at rate
$q=2$ K/min for silica and ethanol, shown as $T_f/T_g$ versus $T/T_g$.
These two systems are chosen as representative limiting cases: silica is a
canonical strong glass former, whereas ethanol is a fragile molecular glass
former. The diagonal line denotes equilibrium, $T_f=T$. The results are obtained
by integrating $dT_f/dt=[T(t)-T_f(t)]/\tau(T)$ with
$\tau(T)=\tau_g\,10^{m(T_g/T-1)}$ and $\tau_g=100$ s.
The parameters used are $T_g\simeq1475$ K and $m\simeq20$ for silica, and
$T_g\simeq97$ K and $m\simeq70$ for ethanol, taken from standard compilations
of glass-forming liquids and fragility data \cite{angell1995,ediger1996,ngai2011}.
Silica remains close to equilibrium over a broad temperature range, whereas
ethanol departs earlier and approaches a nearly constant fictive temperature,
signaling structural arrest.
}
\label{fig:Tf_compare}
\end{figure}

Figure~\ref{fig:Tf_compare} shows the resulting fictive-temperature trajectories.
At high reduced temperature, both systems remain close to equilibrium and follow
the line $T_f=T$. Upon cooling, the relaxation time increases and the system can
no longer equilibrate on the cooling timescale. The departure of $T_f$ from $T$
therefore marks the onset of nonequilibrium structural arrest.

The contrast between the two systems is direct. Silica, with low fragility,
exhibits a gradual departure from equilibrium and remains close to the diagonal
over a broad temperature interval. Ethanol, with much larger fragility, shows an
earlier and sharper departure. At lower temperature, the ethanol curve approaches
a nearly constant fictive temperature, corresponding to the regime
$\tau(T)\gg 1/q$. This flattening is the dynamical signature of frozen
configurational enthalpy.

The numerical comparison therefore illustrates the main physical conclusion:
fragility controls how rapidly the enthalpy relaxation spectrum shifts to long
times during cooling. A fragile liquid loses equilibrium over a narrower
temperature interval and develops a larger fictive-temperature lag, while a
strong liquid follows the bath temperature more closely.

\begin{table}[t]
\centering
\caption{Material parameters used in the numerical calculation of fictive-temperature trajectories. 
The values of $T_g$ and $m$ are representative literature values for silica and ethanol, 
while $\tau_g=100$ s is the conventional calorimetric definition of the glass transition.}
\begin{tabular}{lccc}
\toprule
Material & $T_g$ (K) & Fragility $m$ & $\tau_g$ (s) \\
\midrule
Silica (SiO$_2$) & 1475 & $\sim 20$ & 100 \\
Ethanol & 97 & $\sim 70$ & 100 \\
\bottomrule
\end{tabular}
\label{tab:parameters}
\end{table}

The material parameters employed in the present calculations are representative
literature values widely used in discussions of glass-forming liquids and
fragility. The fragility parameters and glass-transition temperatures for silica
and ethanol were taken from standard compilations and reviews of glass-forming
systems \cite{angell1995,ediger1996,ngai2011}. The purpose of the calculation is
not to reproduce a specific calorimetric experiment quantitatively, but rather to
illustrate how differences in fragility modify the fictive-temperature trajectory
within the present theoretical framework.

%
% ========================  Figure 2  ========================

% ========================= Figure 2 ends =================================

Figure~2 provides a direct visualization of the freezing-out process in
terms of the fictive temperature. At high temperatures, both systems follow
the equilibrium line $T_f = T$, indicating that structural relaxation is
fast compared to the cooling rate. As the temperature approaches $T_g$,
the relaxation time increases rapidly, and the system can no longer
maintain equilibrium. This leads to a departure of $T_f$ from $T$, marking
the onset of nonequilibrium behavior. The effect is markedly different for
strong and fragile liquids: silica, with low fragility, exhibits a gradual
departure and remains close to equilibrium over a wide temperature range,
whereas ethanol, with high fragility, shows an early and pronounced
departure, followed by a rapid saturation of $T_f$. The flattening of the
$T_f$ curve at low temperatures corresponds to the regime where the
structural relaxation time greatly exceeds the inverse cooling rate,
$\tau(T) \gg 1/q$, and the configurational degrees of freedom are effectively
frozen. This provides a clear dynamical interpretation of the frozen
fictive temperature in terms of the arrest of enthalpy fluctuations.

The present formulation also complements recent fluctuation-based descriptions
of glassy relaxation and nonequilibrium thermodynamics \cite{BagchiPCCP2026}.
%
% ============== Section 7 -- on aging  ==========================
%
\section{Connection to RFOT and Energy-Landscape Descriptions}
The present formulation also connects naturally with RFOT-based descriptions of
aging in structural glasses. In RFOT theory, relaxation proceeds through activated
rearrangements among metastable amorphous states, with the relaxation time controlled
by configurational entropy and mismatch penalties
\cite{Kirkpatrick1989,WolynesReview}. 

Extensions of RFOT to aging emphasize that a nonequilibrium glass is characterized
not only by the bath temperature, but also by a history-dependent structural state,
often represented through a fictive temperature. Within this framework, Wolynes
and co-workers developed a local energy-landscape-library picture in which a
rearranging region explores alternative amorphous states while paying a mismatch
free-energy cost to its frozen surroundings. As the region grows, the number of
accessible amorphous states increases, providing a microscopic basis for aging,
rejuvenation, and the evolution of local structural stability. In the aging theory,
the evolving local energy may be mapped onto an effective temperature closely
related to the fictive temperature, and the RFOT treatment also connects the
nonlinearity of aging to kinetic fragility. Subsequent developments incorporated
mobility generation and transport, dynamical heterogeneity, and spatially varying
fictive-temperature and mobility fields
\cite{LubchenkoWolynes2004,Wolynes2009,WisitsorasakWolynes2013,
WisitsorasakWolynes2014,LubchenkoWolynes2017}.

The perspective developed here is complementary. Instead of starting from an
explicit landscape or mobility-field description, we formulate the problem in
terms of the configurational enthalpy--enthalpy time-correlation function. The
same correlation function determines the configurational part of the
frequency-dependent heat capacity and the time evolution of the fictive
temperature. Thus, the slow descent through the energy landscape during aging
may be viewed, in the present formulation, as the progressive relaxation and
spectral shift of configurational enthalpy fluctuations.

This connection suggests a direct bridge between RFOT aging theory and dynamic
calorimetry. RFOT provides a physical picture of activated rearrangements and
history-dependent stability, while the present theory provides an experimentally
accessible representation in terms of $C_p^*(\omega)$, enthalpy fluctuations, and
fictive-temperature dynamics. Fragility then enters through the temperature
dependence of the structural relaxation clock common to both descriptions.

A more detailed incorporation of RFOT aging expressions for the relaxation time,
including explicit configurational-entropy and spatial-heterogeneity effects,
will be considered elsewhere.

A particularly useful result of the RFOT aging theory is its connection between
the kinetic fragility and the nonlinearity parameter of the
Narayanaswamy--Moynihan--Tool description. Lubchenko and Wolynes showed that,
within the approximations of the RFOT treatment, these quantities are related
approximately by
\begin{equation}
m \simeq \frac{19}{x},
\end{equation}
where $x$ is the phenomenological nonlinearity parameter describing the relative
sensitivity of the structural relaxation time to the bath temperature and to
the structural state of the glass. Thus, more fragile liquids are expected to
have smaller values of $x$, and hence a stronger dependence of the relaxation
clock on the nonequilibrium structural state. This result provides an important
connection between fragility and nonlinear aging within RFOT and is
complementary to the present fluctuation-based description, in which fragility
enters through the temperature dependence of the structural relaxation time
governing the evolution of the configurational enthalpy and the fictive
temperature
\cite{LubchenkoWolynes2004,Wolynes2009}.

% ==========================  8 =========================

\section{Conclusion}

In this work, we have developed a fluctuation-based formulation that connects the
fragility parameter \(m\) with the time evolution of the fictive temperature
\(T_f(t)\) during cooling. The connection is most direct and physically
controlled in the regime where the system begins to fall out of equilibrium,
that is, in the initial development of the lag between \(T_f(t)\) and the bath
temperature \(T(t)\). This is the regime in which the equilibrium or
metastable-supercooled-liquid relaxation spectrum near \(T_g\) still controls
the loss of configurational equilibration.

The central starting point is a linear-response formulation in which the
delayed configurational enthalpy response is governed by the configurational
enthalpy--enthalpy time-correlation function. This same correlation function
determines the configurational part of the frequency-dependent heat capacity.
The fictive temperature is then obtained by mapping the nonequilibrium
configurational enthalpy onto the corresponding equilibrium configurational
enthalpy. Thus the formulation naturally leads to a memory equation for
\(T_f(t)\), with the same structural relaxation kernel appearing in both
dynamic calorimetry and fictive-temperature evolution.

A key point is that the fragility parameter \(m\) is defined from the
temperature dependence of the equilibrium structural relaxation time as
\(T_g\) is approached from above. In the present theory, \(m\) therefore enters
through the temperature dependence of the structural clock that controls the
accumulation of reduced time during cooling. This clock determines when and how
rapidly the fictive temperature begins to depart from the bath temperature.
Below \(T_g\), however, the system is no longer in equilibrium at the bath
temperature, and a quantitative description of the subsequent aging or of the
final frozen value of \(T_f\) requires a nonequilibrium continuation of the
relaxation kernel, for example of Tool--Narayanaswamy--Moynihan type.

Two principal results emerge from this formulation. First, we obtain a
time-domain relation between the growth of the fictive-temperature lag and the
structural relaxation kernel. In the Debye limit this relation reduces to
\begin{equation}
\frac{dT_f}{dt}
=
\frac{T-T_f}{\tau(T,T_f)} ,
\end{equation}
where, near the onset of nonequilibrium behavior, the relaxation time may be
approximated by its equilibrium supercooled-liquid form,
\begin{equation}
\tau(T)\simeq
\tau_g\,10^{\,m\left(\frac{T_g}{T}-1\right)} .
\end{equation}
This result shows explicitly how fragility controls the rate at which
\(T_f(t)\) begins to depart from \(T(t)\) during cooling.

Second, we show that the temperature dependence of the calorimetric loss
peak, \(C_p''(\omega)\), is governed by the same structural clock. Under
thermorheological simplicity,
\begin{equation}
\omega_p(T) \propto
10^{-\,m\left(\frac{T_g}{T}-1\right)} ,
\end{equation}
so that the slope of \(\log_{10}\omega_p\) versus \(T_g/T\) provides the same
fragility parameter that characterizes the temperature dependence of the
structural relaxation time. This establishes a direct connection between
Angell fragility, the spectral shift of configurational enthalpy fluctuations,
and the onset of fictive-temperature lag.

The theory therefore provides a unified physical picture: fragility measures
the rate at which the structural relaxation clock slows down upon cooling,
which in turn controls the accumulation of reduced time and the progressive
freezing of the configurational degrees of freedom. Within this framework, the
shape of the relaxation function is determined by the configurational
enthalpy-relaxation spectrum, while fragility governs how rapidly that spectrum
shifts to longer times as the temperature is lowered.

The resulting trends are consistent with the standard phenomenology of strong
and fragile glass formers. Fragile liquids exhibit a sharper departure from
equilibrium and develop a pronounced lag between \(T_f\) and \(T\), whereas
strong glass formers remain closer to equilibrium over a broader temperature
range. Within the approximate freezing criterion used here, the frozen fictive
temperature also acquires a logarithmic cooling-rate dependence whose slope is
controlled by fragility. Similarly, the temperature shift of the calorimetric
loss peak provides an independent manifestation of the same structural clock.

\textit{A central result of the present work is that fragility can be expressed
in terms of the temperature evolution of the configurational enthalpy-relaxation
spectrum.} This provides a time-correlation-function-based interpretation of
fragility that complements its conventional definition in terms of the
temperature dependence of structural relaxation times.

It is useful to emphasize the role of linear response in the present
formulation. We have not treated \(T_f(t)-T(t)\) as an externally imposed
Kubo field. Instead, linear response enters through the experimentally
measurable frequency-dependent heat capacity. The configurational part
of \(C_p^{*}(\omega)\) determines the delayed enthalpy response to a
prescribed thermal history, and the fictive temperature is then obtained
by mapping this nonequilibrium enthalpy onto the corresponding
equilibrium configurational enthalpy. Thus the response function is used
in its calorimetric sense, while \(T_f(t)-T(t)\) appears as the thermodynamic
measure of the resulting enthalpy lag.

The present approach is also connected to earlier work in which the
frequency-dependent heat capacity was obtained from microscopic theories
of enthalpy fluctuations, including mode-coupling-type and memory-function
descriptions. In this sense, the present theory may be viewed as using
\(C_p^{*}(\omega)\) as the bridge between microscopic enthalpy dynamics and
the macroscopic fictive-temperature description.

Several limitations of the present formulation should be kept in mind.
First, the derivation assumes a near-linear-response regime, so that the
mapping between the configurational enthalpy lag and \(T_f(t)-T(t)\) can be
linearized. Second, the use of reduced time assumes thermorheological
simplicity: the relaxation spectrum may shift strongly with temperature,
but its normalized shape is assumed not to change appreciably. This
approximation can fail in systems with pronounced dynamic heterogeneity,
multiple relaxation channels, strong secondary relaxations,
crystallization, phase separation, or large structural changes during
aging. Third, the Debye-limit equation used for numerical illustration is
only the simplest limit of the more general convolution relation;
quantitative description of real materials may require stretched-exponential
or distributed relaxation kernels.

The numerical figures presented here should therefore be interpreted as
illustrations of the central physical mechanism rather than as detailed
fits to particular calorimetric protocols. More quantitative applications
would require experimentally measured \(C_p^{*}(\omega,T)\) over the relevant
temperature range, including the temperature dependence of \(\Delta C_p\),
possible changes in the spectral shape, and, when necessary, explicit
dependence of the relaxation time on both \(T\) and \(T_f\) in the spirit of
TNM theory.

The present work thus provides a direct bridge between fragility,
configurational enthalpy relaxation, and nonequilibrium fictive-temperature
dynamics. The formulation opens the way to further developments, including
extensions beyond the Debye limit, incorporation of nonexponential relaxation,
and application to specific systems where the configurational heat capacity
and relaxation spectra can be independently measured.

%=====================================================

%===============================================
%
\end{document}